\documentclass[runningheads]{llncs}
\usepackage{graphicx}
\usepackage{amsmath}
\usepackage{bbm}
\usepackage{multirow}
\usepackage[font=small,labelfont=bf]{caption}
\begin{document}
\title{Extending LOUPE for K-space Under-sampling Pattern Optimization in Multi-coil MRI}
\titlerunning{Optimized k-space sampling using extended LOUPE}
%
\author{Jinwei Zhang\inst{1,3} \and
Hang Zhang\inst{2,3} \and
Alan Wang\inst{2} \and
Qihao Zhang\inst{1,3} \and
Mert Sabuncu\inst{1,2,3} \and
Pascal Spincemaille\inst{3} \and
Thanh D. Nguyen\inst{3} \and
Yi Wang\inst{1,3}}

\authorrunning{J. Zhang et al.}
%
\institute{Department of Biomedical Engineering, Cornell University, Ithaca, NY, USA \and
Department of Electrical and Computer Engineering, Cornell University, Ithaca, NY, USA \and
Department of Radiology, Weill Medical College of Cornell University, New York, NY, USA}
\maketitle              
\begin{abstract}
The previously established LOUPE (Learning-based Optimization of the Under-sampling Pattern) framework for optimizing the k-space sampling pattern in MRI was extended in three folds: firstly, fully sampled multi-coil k-space data from the scanner, rather than simulated k-space data from magnitude MR images in LOUPE, was retrospectively under-sampled to optimize the under-sampling pattern of in-vivo k-space data; secondly, binary stochastic k-space sampling, rather than approximate stochastic k-space sampling of LOUPE during training, was applied together with a straight-through (ST) estimator to estimate the gradient of the threshold operation in a neural network; thirdly, modified unrolled optimization network, rather than modified U-Net in LOUPE, was used as the reconstruction network in order to reconstruct multi-coil data properly and reduce the dependency on training data. Experimental results show that when dealing with the in-vivo k-space data, unrolled optimization network with binary under-sampling block and ST estimator had better reconstruction performance compared to the ones with either U-Net reconstruction network or approximate sampling pattern optimization network, and once trained, the learned optimal sampling pattern worked better than the hand-crafted variable density sampling pattern when deployed with other conventional reconstruction methods. 

\end{abstract}
\section{Introduction}
Parallel imaging (PI) \cite{griswold2002generalized,pruessmann1999sense} and Compressed Sensing MRI (CS-MRI) \cite{lustig2007sparse} are widely used technique for acquiring and reconstructing  under-sampled k-space data thereby shortening scanning times in MRI. CS-MRI is a computational technique that suppresses incoherent noise-like artifacts introduced by random under-sampling, often via a regularized regression strategy. Combining CS-MRI with PI promises to make MRI much more accessible and affordable. Therefore, this has been an intense area of research in the past decade \cite{feng2014golden,murphy2012fast,vasanawala2011practical}. One major task in PI CS-MRI is designing a random under-sampling pattern, conventionally controlled by a variable-density probabilistic density function (PDF). However, the design of the ‘optimal’ under-sampling pattern remains an open problem for which heuristic solutions have been proposed. For example, \cite{knoll2011adapted} generated the sampling pattern based on the power spectrum of an existing reference dataset; \cite{haldar2019oedipus} combined experimental design with the constrained Cramer-Rao bound to generate the context-specific sampling pattern; \cite{gozcu2018learning} designed a parameter-free greedy pattern selection method to find a sampling pattern that performed well on average for the MRI data in a training set.

Recently, with the success of learning based k-space reconstruction methods \cite{aggarwal2018modl,hammernik2018learning,schlemper2017deep,zhang2020fidelity}, a data-driven machine learning based approach called LOUPE \cite{bahadir2019learning} was proposed as a principled and practical solution for optimizing the under-sampling pattern in CS-MRI. In LOUPE, fully sampled k-space data was simulated from magnitude MR images and retrospective under-sampling was deployed on the simulated k-space data. A sampling pattern optimization network and a modified U-Net \cite{ronneberger2015u} as the under-sampled image reconstruction network were trained together in LOUPE to optimize both the k-space under-sampling pattern and reconstruction process. In the sampling pattern optimization network, one sigmoid operation was used to map the learnable weights into probability values, and a second sigmoid operation was used to approximate the non-differentiable step function for stochastic sampling, as the gradient needed to be back-propagated through such layer to update the learnable weights. After training, both optimal sampling pattern and reconstruction network were obtained. For a detailed description of LOUPE we refer the reader to \cite{bahadir2019learning}.

In this work, we extended LOUPE in three ways. Firstly, in-house multi-coil in-vivo fully sampled T2-weighted k-space data from MR scanner was used to learn the optimal sampling pattern and reconstruction network. Secondly, modified U-Net \cite{ronneberger2015u} as the reconstruction network in LOUPE was extended to a modified unrolled reconstruction network with learned regularization term in order to reconstruct multi-coil data in PI with proper data consistency and reduce the dependency on training data when training cases were scarce. Thirdly, approximate stochastic sampling layer was replaced by a binary stochastic sampling layer with Straight-Through (ST) estimator \cite{bengio2013estimating}, which was used to avoid zero gradients when back-propagating to this layer. Fully sampled data was acquired in healthy subjects. Under-sampled data was generated by retrospective under-sampling using various sampling patterns. Reconstructions were performed using different methods and compared.

\section{Method}
In PI CS-MRI, given an under-sampling pattern and the corresponding acquired k-space data, a reconstructed image $\hat{x}$ is obtained via minimizing the following objective function:
\begin{equation}\label{eqn::csmri}
    \hat{x}=\arg\min_x \Sigma_j^{N_c}\| UFS_jx - b_j\|_2^2 + R(x),
\end{equation}
where $x$ the MR image to reconstruct, $S_j$ the coil sensitivity map of $j$-th coil, $N_c$ the number of receiver coils, $F$ the Fourier transform, $U$ the k-space under-sampling pattern, and $b_j$ the acquired under-sampled k-space data of the $j$-th coil. $R(x)$ is a regularization term, such as Total Variation (TV) \cite{osher2005iterative} or wavelet \cite{donoho1995nonlinear}. The minimization in Eq. \ref{eqn::csmri} is performed using iterative solvers, such as the Quasi-Newton method \cite{dennis1977Quasi}, the alternating direction method of multipliers (ADMM) \cite{boyd2011distributed} or the primal-dual method \cite{chambolle2011first}. Eq. \ref{eqn::csmri} can also be mimicked by learning a parameterized mapping such as neural network from input $\{b_j\}$ to output $\hat{x}$. We denote the mapping $\{b_j\} \to \hat{x}$ using either iterative solvers or deep neural networks as $\hat{x} = \mathcal{A}(\{b_j\})$.

Our goal is to obtain an optimal under-sampling pattern $\hat{U}$ for a fixed under-sampling ratio $\gamma$ from $N$ fully sampled data through retrospective under-sampling. The mathematical formulation of this problem is:
\begin{equation}\label{eqn::loupe}
    \min_{U} \frac{1}{N} \Sigma_{i=1}^N L(x_i^*, \hat{x}_i(U)), \ 
    \text{subject to} \ U \in \Omega, \
    \hat{x_i}(U) = \mathcal{A}(\{Ub^*_{ij}\}),
\end{equation}
where $x_i^*$ the $i$-th MR image reconstructed by direct inverse Fourier transform from fully sampled k-space data $\{b^*_{ij}\}$, $L(\cdot, \cdot)$ the loss function to measure the similarity between reconstructed image $\hat{x_i}(U)$ and fully sampled label $x_i^*$, $\Omega$ the constraint set of $U$ to define how $U$ is generated with a fixed under-sampling ratio $\gamma$. The bilevel optimization problem \cite{colson2007overview} of Eq. (\ref{eqn::loupe}) was solved in LOUPE \cite{bahadir2019learning} via jointly optimizing a modified U-Net \cite{ronneberger2015u} as $\mathcal{A}$ and an approximate stochastic sampling process as $\Omega$ on a large volume of simulated k-space data from magnitude MR images. However, for in-vivo k-space data with multi-coil acquisition as in PI, both U-Net architecture for reconstruction and approximate stochastic sampling for pattern generation could be sub-optimal. Specifically, due to limited training size of in-vivo data and no k-space consistency imposed in U-Net, inferior reconstructions could happen in test and even training datasets. And the approximate stochastic sampling process generated fractional rather than 0-1 binary patterns during training, which might not work well during test as binary patterns should be used for realistic k-space sampling.
In view of the above, we extend and improve LOUPE in terms of both reconstruction mapping $\mathcal{A}$ and sampling pattern's generating process $\Omega$ when working on in-vivo multi-coil k-space data in this work.

\subsection{Unrolled Reconstruction Network}
A modified residual U-Net \cite{ronneberger2015u} was used as the reconstruction network in LOUPE \cite{bahadir2019learning} to map from the zero-filled k-space reconstruction input to the fully-sampled k-space reconstruction output. U-Net works fine with simulated k-space reconstruction when enough training data of magnitude MR images are given, but as for in-vivo multi-coil k-space data, training cases are usually scarce, since fully-sampled scans are time consuming and as a result, only a few fully-sampled cases can be acquired. 

To reduce the dependency on training dataset and improve the data consistency of deep learning reconstructed images, combining neural network block for the regularization term in Eq. \ref{eqn::csmri} with iterative optimization scheme to solve Eq. \ref{eqn::csmri} has been explored in recent years \cite{aggarwal2018modl,hammernik2018learning,schlemper2017deep}, which are called "unrolled optimization/reconstruction networks" in general. Prior works showed that such unrolled networks performed well for multi-coil k-space reconstruction task by means of inserting measured k-space data into the network architecture to solve Eq. \ref{eqn::csmri} with a learning-based regularization. In light of the success of such unrolled reconstruction networks, we apply a modified MoDL \cite{aggarwal2018modl} as the reconstruction network in this work. MoDL unrolled the quasi-Newton optimization scheme to solve Eq. \ref{eqn::csmri} with a neural network based denoiser as the $L_2$ regularization term $R(x)$, and conjugate gradient (CG) descent block was applied in MoDL architecture to solve $L_2$ regularized problem. Besides, we will show that such unrolled network architecture also works as the skip connections for sampling pattern weights' updating as the generated pattern is connected to each intermediate CG block to perform $L_2$ regularized data consistency (Fig. \ref{fig1}).

\subsection{ST Estimator for Binary Pattern}
 In LOUPE \cite{bahadir2019learning}, a probabilistic pattern $P_m$ was defined as $P_m = \frac{1}{1 + e^{-a \cdot w_m}}$ with hyper-parameter $a$ and trainable weights $w_m$. The binary k-space sampling pattern $U$ was assumed to follow a Bernoulli distribution $Ber(P_m)$ independently on each k-space location. $U$ was generated from $P_m$ as $U =  \textbf{1}_{z < P_m}$, where $z \sim U[0, 1]^{\dim(P_m)}$ and $\textbf{1}_x$ the pointwise indicator function on the truth values of $x$. However, indicator function $\textbf{1}_x$ has zero gradient almost everywhere when back-propagating through it. LOUPE addressed this issue by approximating $\textbf{1}_{z < P_m}$ using another sigmoid function: $U \approx  \frac{1}{1 + e^{-b \cdot (P_m - z)}}$ with hyper-parameter $b$. 
 
 Although the gradient issue was solved in LOUPE, $U$ was approximated as a fraction between $[0, 1]$ on each k-space location instead of the binary pattern deployed in both test phase and realistic MR scan. As a result, binary sampling patterns generated in test phase could yield inferior performance due to such mismatch with training phase. To address this issue, binary patterns are also needed during training phase, at the same time gradient back-propagating through binary sampling layer should be properly handled. Such binary pattern generation layer can be regarded as the layer with stochastic neurons in deep learning, and several methods have been proposed to address its back-propagation \cite{bengio2013estimating,hinton2012neural}. Here we use straight through (ST) estimator \cite{bengio2013estimating} in the stochastic sampling layer to generate binary pattern $U$ meanwhile addressing the zero gradient issue during back-propagation. Based on one variant of ST estimator, $U$ is set as $\textbf{1}_{z < P_m}$ during forward pass. When back-propagating through the stochastic sampling layer, an ST estimator replaces the derivative factor $\frac{d\textbf{1}_{z < P_m}}{dw_m} = 0$ with the following:
\begin{align}
    \frac{d\textbf{1}_{z < P_m}}{dw_m} = \frac{d P_m}{dw_m}.
\end{align}
In other words, indicator function in the stochastic layer is applied at forward pass but treated as identity function during back-propagation. This ST estimator allows the network to make a yes/no decision, allowing it to picking up the top $\gamma$ fraction  of k-space locations most important for our task.

\subsection{Network Architecture}
\begin{figure}
\vspace{-2mm}
\includegraphics[width=\textwidth]{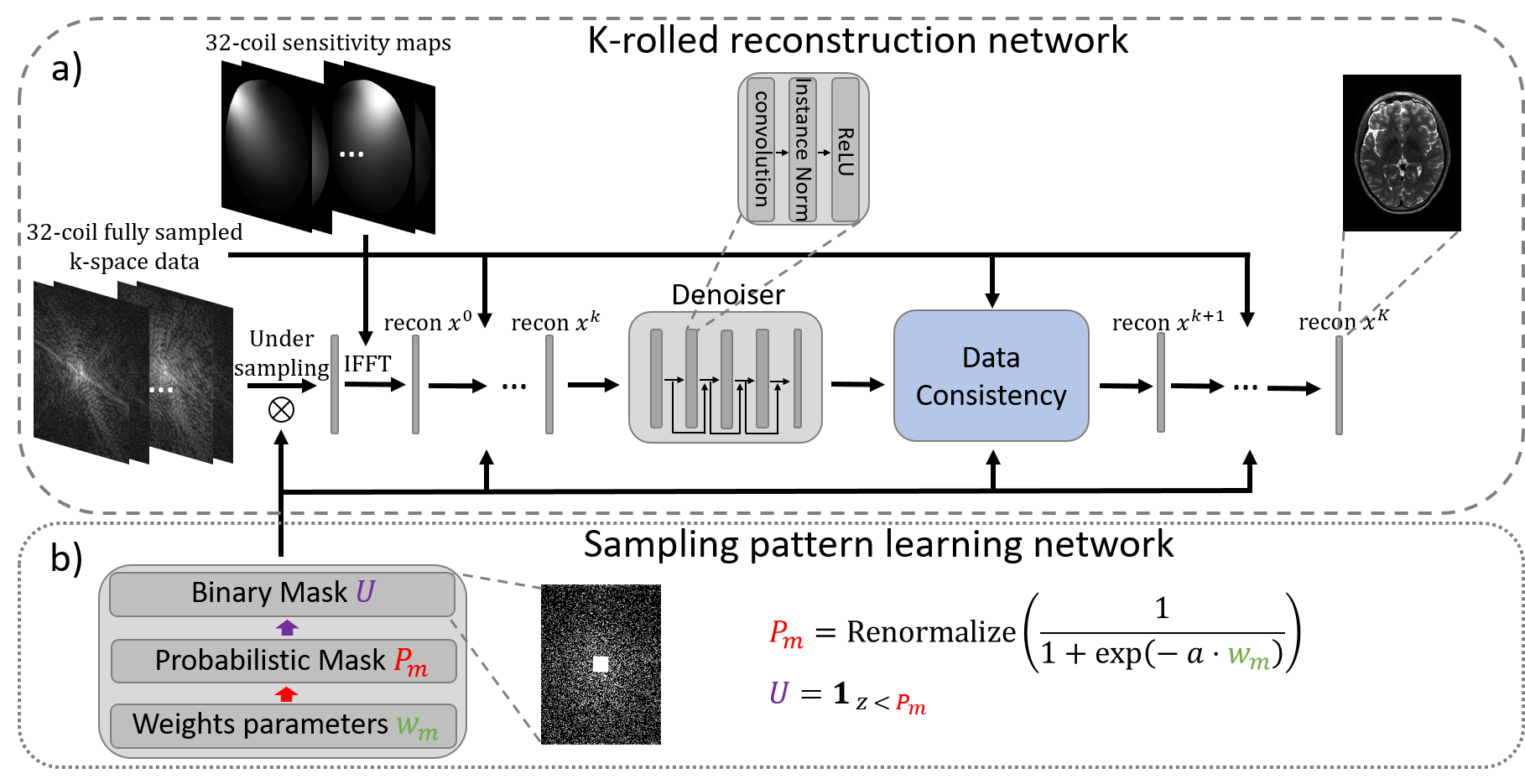}
\centering
\caption{Proposed network architecture consisting of a sampling pattern learning network and a K-rolled reconstruction network.} \label{fig1}
\end{figure}

Fig. \ref{fig1} shows the proposed network architecture consisting of two sub-networks: one unrolled reconstruction network and one sampling pattern learning network. 

In the sampling pattern learning network (Fig. \ref{fig1}(b)), Renormalize($\cdot$) is a linear scaling operation to make sure the mean value of probabilistic pattern is equal to the desired under-sampling ratio $\gamma$. The binary pattern $U$ is sampled at every forward pass in the network and once generated, it is used to retrospectively under-sample the fully sampled multi-coil k-space data. 

The deep quasi-Newton network (MoDL \cite{aggarwal2018modl}) as the unrolled reconstruction network architecture is illustrated in Fig. \ref{fig1}(a). In deep quasi-Newton, Denoiser + Data consistency blocks are replicated $K$ times to mimic $K$ quasi-Newton outer loops of solving Eq. \ref{eqn::csmri} in which a neural network denoiser for $R(x)$ is applied. Five convolutional layers with skip connection \cite{he2016deep} and instance normalization \cite{ulyanov2016instance} are used as the denoiser and the weights are shared among blocks. The binary pattern $U$ is used to generate zero-filled reconstruction $x^0$ as the input of reconstruction network and also connected to all the data consistency sub-blocks to deploy regularized optimization, which also works as the skip connection to benefit the training of pattern weights $w_m$. 


\section{Experiments}
\subsection{Dataset and Implementations}

\textbf{Data acquisition and processing.}
Fully sampled k-space data were acquired in 6 healthy subjects ($5$ males and $1$ female; age: $30 \pm 6.6$ years) using a sagittal T2-weighted variable flip angle 3D fast spin echo sequence on a 3T GE scanner with a 32-channel head coil. Imaging parameters were: $256\times256\times192$ imaging matrix, $1mm^3$ isotropic resolution. Coil sensitivity maps of each axial slice were calculated with ESPIRiT \cite{uecker2014espirit} using a $25\times25\times32$ auto-calibration k-space region. From the fully sampled data, a combined single coil image using the same coil sensitivity maps was computed to provide the ground truth label for both sampling pattern learning and reconstruction performance comparison. The central 100 slices of each subject were extracted for the training (300 slices), validation (100 slices) and test (200 slices) dataset. In addition, k-space under-sampling was performed retrospectively in the ky-kz plane for all the following experiments.
\\ 

\begin{figure}[t!]
\centering
\includegraphics[width=0.9\textwidth]{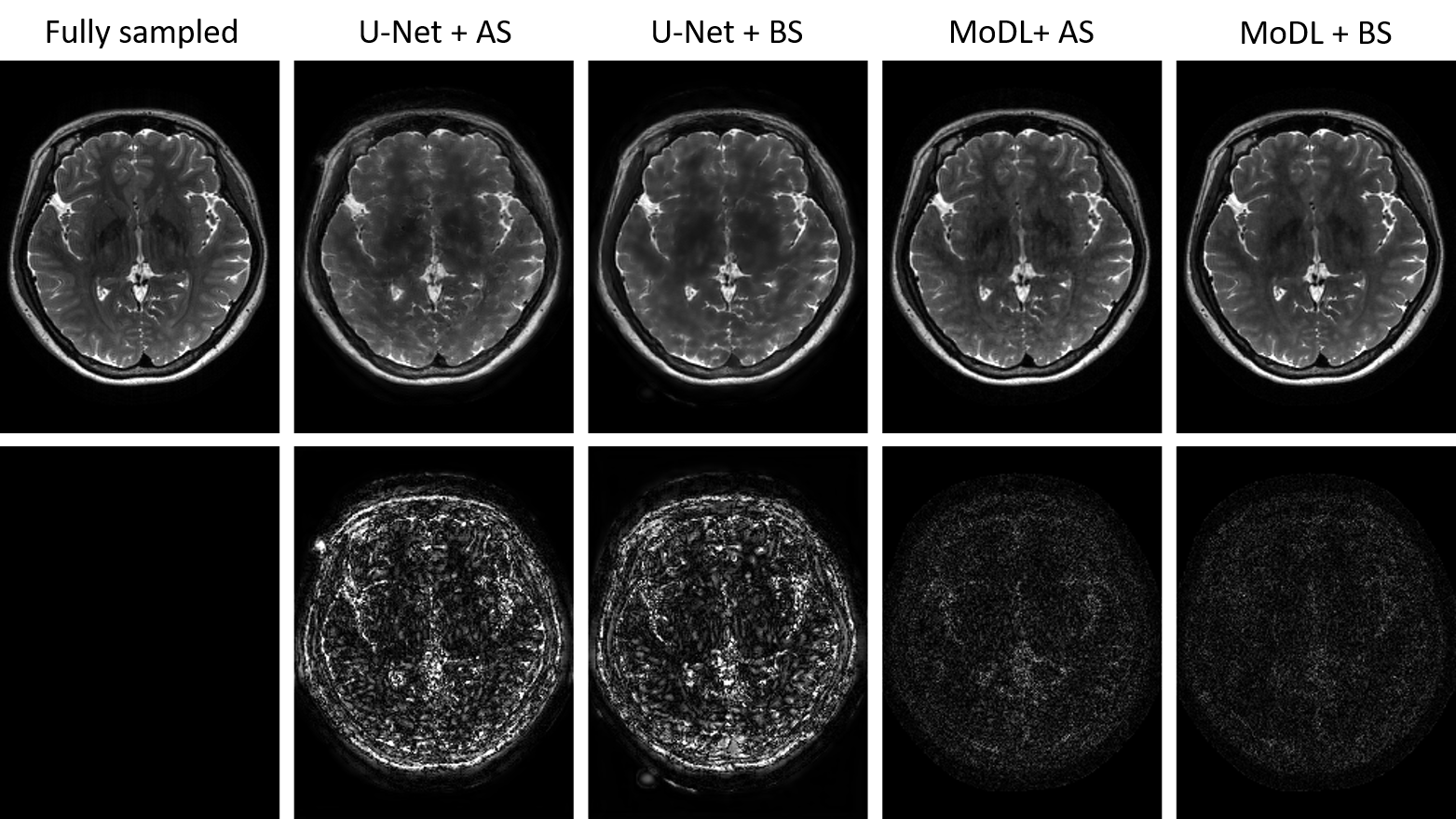}
\caption{Reconstruction results on one test slice by four combinations of reconstruction network and sampling pattern optimization network with 10\% under-sampling ratio. First row: reconstruction results; second row: $5 \times$ absolute error maps (window level: [0, 0.5]). MoDL + BS equipped with ST estimator had the best performance.} \label{fig2}
\end{figure}

\noindent \textbf{Training parameters.} In the sampling pattern learning network, $w_m$ were initialized randomly, the slope factor $a = 0.25$ and the under-sampling ratio $\gamma = 10\%$. The central $25 \times 25$ k-space region remained fully sampled for each pattern. For the baseline LOUPE, a second slope factor $b = 12$ was used to approximate the binary sampling. The sampling pattern learning networks using binary sampling with ST estimator and approximated sampling were denoted as BS (binary sampling) and AS (approximated sampling) in the following experiments. In the unrolled reconstruction network, $K=5$ replicated blocks were applied and the denoiser was initialized randomly. For the baseline LOUPE, a residual U-Net was applied. All of the learnable parameters in Fig. \ref{fig1} were trained simultaneously using the loss function: $\frac{1}{N}\Sigma_{i=1}^{N}\Sigma_{k=1}^K \| x_i^k - x_i^*\|_1$, 
where $x_i^*$ the $i$-th ground truth label in the training dataset, $x_i^k$ the $k$-th intermediate reconstruction ($K=1$ in U-Net). Stochastic optimization with batch size $1$ and Adam optimizer (initial learning rate: $10^{-3}$) \cite{kingma2014adam} was used to minimize the loss function. The number of epochs was $200$. The whole training and inference procedures were implemented in PyTorch with Python version 3.7.3 on an RTX 2080Ti GPU.








\subsection{Comparison with LOUPE}
Fig. \ref{fig2} shows the reconstruction results from one of the test subjects to demonstrate the performance improvement of the extended LOUPE over vanilla LOUPE. Four combinations of reconstruction network and sampling pattern optimization network were tested and compared. Binary sampling patterns were generated during test phase. From Fig. \ref{fig2}, MoDL provided better reconstruction results compared to U-Net, while for both U-Net and MoDL reconstruction networks, BS (binary sampling) gave less noisy reconstructions than AS (approximate sampling) during test phase. Quantitative comparisons in terms of PSNR (peak signal-to-noise ratio) and SSIM (structural similarity index measure \cite{wang2004image}) are shown in Table \ref{tab1a}, where MoDL + BS had the best performance.


\begin{table}
    \centering
    \caption{Quantitative results of section 3.2} \label{tab1a}
    \begin{tabular}{c|c|c}
        \hline
         & PSNR (dB) & SSIM \\
        \hline
        U-Net+AS & 32.5 $\pm$ 1.0 & 0.885 $\pm$ 0.016 \\
        
        U-Net+BS & 33.0 $\pm$ 0.6 & 0.898 $\pm$ 0.012 \\
        
        MoDL+AS  & 41.3 $\pm$ 1.2  & 0.963 $\pm$ 0.015 \\
        
        MoDL+BS  & $\bf{42.6} \ \pm$ 1.1 & $\bf{0.968} \ \pm$ 0.012 \\
        \hline
    \end{tabular}%
 \vspace{-10mm}
 \end{table}

 \begin{table}
    \centering
    \caption{Quantitative results of section 3.3} \label{tab1b}
        \begin{tabular}{ c|c|c|c }
            \hline
            Pattern & Method & PSNR (dB) & SSIM \\
            \hline
            \multirow{3}{*}{VD} & ESPIRiT & 37.5 $\pm$ 1.0 & 0.920 $\pm$ 0.016 \\
             & TGV & 40.1 $\pm$ 0.9 & 0.952 $\pm$ 0.014 \\
             & MoDL & 40.4 $\pm$ 0.9 & 0.963 $\pm$ 0.010 \\
            \hline
            \multirow{3}{*}{Learned} & ESPIRiT & $\bf{39.5} \ \pm$ 1.1 & $\bf{0.932} \ \pm$ 0.018 \\
             & TGV & $\bf{42.5} \ \pm$ 1.1 & $\bf{0.959} \ \pm$ 0.016 \\
             & MoDL & $\bf{42.6} \ \pm$ 1.1 & $\bf{0.968} \ \pm$ 0.012 \\
            \hline
        \end{tabular}
\vspace{-5mm}
\end{table}

 \begin{figure}[h!]
\centering
\includegraphics[width=0.9\textwidth]{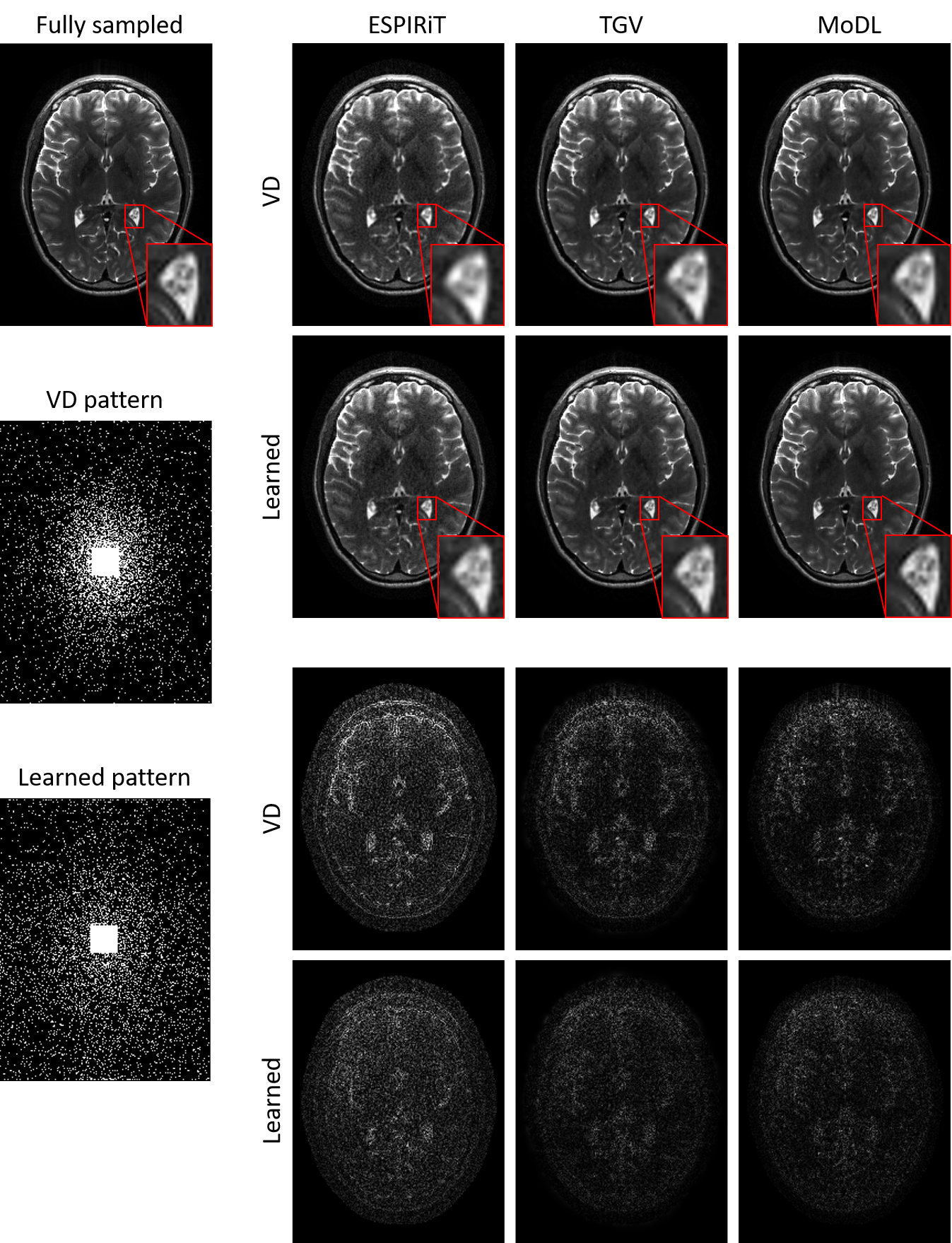}
\caption{Reconstruction results on another test slice using VD and learned sampling patterns with three different reconstruction methods. First two rows: reconstruction results; last two rows: corresponding 5$\times$ absolute error maps (window level:[0, 0.5]). For each reconstruction method, the learned sampling pattern produced lower global errors and sharper structural details than VD sampling pattern.} \label{fig3}
\end{figure}

\subsection{Comparison with other pattern}
To compare the learned sampling pattern ('learned pattern' in Fig. \ref{fig3}, generated from MoDL + BS in section 3.2) with the manually designed one with 10 \% ratio, a variable density (VD) sampling pattern following a probabilistic density function whose formula is a polynomial of the radius in k-space with tunable parameters \cite{uecker2015berkeley} was generated ('VD pattern' in Fig. \ref{fig3}). ESPIRiT \cite{uecker2014espirit} and TGV \cite{knoll2011second} as two representative iterative methods for solving PI CS-MRI were also deployed using both sampling patterns, and the corresponding reconstruction results are shown in Fig. \ref{fig3}. For each reconstruction method, the learned sampling pattern captured better image depictions with lower global errors than VD pattern and the structural details as zoomed in were also sharper with the learned sampling pattern. PSNR and SSIM in Table \ref{tab1b} shows consistently improved performance of the learned sampling pattern over the VD pattern for each reconstruction method.

\section{Conclusions}

In this work, LOUPE for optimizing the k-space sampling pattern in MRI was extended by training on in-vivo multi-coil k-space data and using the unrolled network for under-sampled reconstruction and binary stochastic sampling with ST estimator for sampling pattern optimization. Experimental results show that the extended LOUPE worked better than vanilla LOUPE on in-vivo k-space data and the learned sampling pattern also performed well on other reconstruction methods. Future work includes implementing the learned sampling pattern in the pulse sequence to optimize the k-space data acquisition process prospectively.

%
%
%
\bibliographystyle{splncs04}
\bibliography{samplebibliography}

\end{document}